**Quantifying exchange forces of a spin spiral on the atomic scale**

Nadine Hauptmann[1*], Soumyajyoti Haldar[2], Tzu-Chao Hung[1], Wouter Jolie[1], Mara Gutzeit[2], Daniel Wegner[1], Stefan Heinze[2], Alexander A. Khajetoorians[1]

[1] Institute for Molecules and Materials, Radboud University, 6525 AJ Nijmegen, Netherlands

[2] Institut für Theoretische Physik und Astrophysik, Christian-Albrechts-Universität, Kiel, Germany

[*] Correspondence to: n.hauptmann@science.ru.nl

**The large interest in chiral magnetic structures for realization of nanoscale magnetic storage or logic devices has necessitated methods which can quantify magnetic interactions at the atomic scale. To overcome the limitations of the typically used current-based sensing of atomic-scale exchange interactions, a force-based detection scheme is highly advantageous. Here, we quantify the atomic-scale exchange force field between a ferromagnetic tip and a cycloidal spin spiral using our developed combination of current and exchange force detection. Compared to the surprisingly weak spin polarization, the exchange force field is more sensitive to atomic-scale variations in the magnetization. First-principles calculations reveal that the measured atomic-scale variations in the exchange force originate from different contributions of direct and indirect (Zener) type exchange mechanisms, depending on the chemical tip termination. Our work opens the perspective of quantifying different exchange mechanisms of chiral magnetic structures with atomic-scale precision using 3D magnetic exchange force field measurements.**

Chiral nanoscale spin structures such as magnetic skyrmions[1, 2], as well as individual magnetic adatoms[3], have raised a lot of interest in recent years due to their potential for magnetic storage or logic devices at the atomic scale[4, 5]. This has necessitated methods that can quantify and manipulate magnetic interactions between coupled magnetic atoms with atomic-scale precision. Spin-sensitive approaches based on scanning tunneling microscopy (STM) are sensitive to local variations in the exchange interaction at the atomic scale by detecting inelastic tunneling[6, 7, 8, 9], spin resonance[10, 11], or magnetization between a spin-sensitive tip and a surface[12, 13]. Most notable of these methods is spin-polarized scanning tunneling microscopy (SP-STM), in which the total detected current is sensitive to the relative magnetization between the surface and a magnetized tip[14, 15, 16]. SP-STM has been widely used to characterize non-collinear magnetization textures[17, 18] driven by the Dzyaloshinskii-Moriya interaction (DMI)[19, 20] that occurs in structures with broken inversion symmetry such as surfaces due to spin-orbit coupling. In metals the DMI is mediated by conduction electrons[21]. The Hamiltonian of the DMI between two spins $\boldsymbol{S}_i$ and $\boldsymbol{S}_j$ is $H = -\boldsymbol{D}_{ij} \cdot (\boldsymbol{S}_i \times \boldsymbol{S}_j)$ where $\boldsymbol{D}_{ij}$ is the DM vector the direction of which is given by symmetry[19] and the cross-product implies that canted spin structures are preferred. However, the tunneling current is also sensitive to the tunneling anisotropic magnetoresistance[22], the non-collinear magnetoresistance[23], as well as structural and electronic variations, which are all convoluted in the overall measured signal.

Magnetic exchange force microscopy (MExFM) is a method, based on non-contact atomic force microscopy (nc-AFM), complementary to spin-sensitive tunneling spectroscopy. It directly measures the local exchange force between a magnetic tip and the surface[24, 25]. Force-based detection of magnetic interactions presents numerous advantages over current-based detection. Most prominently, MExFM can address magnetic insulators[24, 26], and it can be combined with simultaneous spin-polarized current detection (SPEX)[27] to disentangle the geometrical structure from electronic properties of magnetic structures[28, 29]. Moreover, magnetic exchange force spectroscopy (MExFS) can directly quantify exchange forces[27, 30], which, combined with first-principles calculations, allows for determining the interplay of various exchange mechanisms[31] as well as the role of chemical functionalization of the tip[32, 33]. Nevertheless, this method has been scarcely applied, in contrast to the non-magnetic 3D chemical force field method that has been widely used to quantify chemical forces between the tip and various surfaces[34, 35] as well as molecules[36]. However, 3D-MExFS contributes to answer open questions in surface magnetism, for example, what the role of spin-orbit coupling on the total force is, in analogy to the modification of the spin-averaged electronic structure known as the tunneling anisotropic magnetoresistance. A further question is if it is possible to detect and quantify various exchange mechanisms, including Zener-type exchange and anisotropic exchange, of chiral magnetic structures with atomic-scale resolution.

Here, we utilize SPEX to investigate a single atomic layer of Mn on W(110), which exhibits a cycloidal spin-spiral ground state driven by the DMI[17]. We demonstrate that the spatially resolved exchange force is much more sensitive to the atomic-scale magnetic exchange force field, including the local atomic sites within the magnetic unit cell, when directly compared to the spin polarization. Utilizing simultaneous MExFS and *I-z* spectroscopy (MExFIS), we map out the evolution of the exchange force field and spin polarization between various non-collinear orientations of the tip magnetization relative to the sample. Using first-principles calculations based on density functional theory, taking into account both the tip and the surface, we understand the experimental findings of the exchange force field, and we quantify the influence of chemically different tip terminations on the overall distance dependence and the various possible exchange pathways. The surprisingly weak spin polarization of the tunneling current is explained by the dominant contribution of low spin-polarized $p_z$-states near the Fermi energy, $E_F$, due to the large exchange splitting of $d$-bands in Mn.

## Results

### Atomic-resolution imaging of the Mn spin spiral.

We sense the atomic-scale spatial variation of the magnetic exchange interaction of the cycloidal spin spiral utilizing constant-height (ch) imaging. Fig. 1(d) sketches the cycloidal spin spiral magnetic ground state along the $[1\bar{1}0]$ direction of the Mn monolayer, which has a periodicity of $\lambda \sim 12$ nm, as shown previously by SP-STM imaging[17]. The angle between the magnetic moments of adjacent Mn rows is about 173° resulting in a locally almost antiferromagnetic rectangular c(2 x 2) unit cell (Fig. 1(a)). Fig. 1 shows simultaneously measured maps of (b) the current $I_T^{ch}$, and (c) the frequency shift $\Delta f^{ch}$ and at a constant height $z_1$ that is determined by opening the feedback loop at height $z_0$, at which constant-current (cc) SP-STM images (Supplementary Fig. 1) were acquired ($V_s$ = -10 mV; $I_T^{cc}$ = -2 nA), and moving the tip by a tip displacement of $\Delta z$ = -0.29 nm. We note that negative $\Delta z$ reflects smaller tip-surface distances and $\Delta z = 0$ nm corresponds to our reference height $z_0$). $I_T^{ch}$ shows a stripe pattern along the [001] direction (Fig. 1(b)), illustrating spin contrast consistent with the spin spiral order reported in Ref. [17]. A long-range modulation of the stripe-pattern contrast is observed with a periodicity of $\Delta x_m$ = (5.70 ± 0.03) nm reflecting half the periodicity ($\lambda/2$) of the spin spiral[17]. We refer to the position with maximum contrast of the stripes as $\varphi = 0°$ and $\varphi = 180°$ (Fig. 1(d)), which correspond to positions with reverse alignments of the out-of-plane magnetic moments. We have carefully acquired and analyzed all data to make sure that the tip has a dominantly magnetic out-of-plane sensitivity (see Supplementary Note 1 and Supplementary Fig. 1 for more details). We therefore can assign the darker and brighter stripes of $I_T^{ch}$ to Mn rows exhibiting an antiparallel (*ap*) alignment at $\varphi = 0°$ and a parallel (*p*) alignment at $\varphi = 180°$ with respect to the out-of-plane projection of the tip magnetization (see Fig. 1(d)). The periodicity of $\varDelta x_1 = (0.43 \pm 0.03)$ nm along the $[1\bar{1}0]$ corresponds to the lattice constant of the magnetic c(2 x 2) unit cell as sketched in Fig. 1(a)[14]. In agreement with previous works[14, 17], the lattice constant along the [001] direction ($\varDelta x_2$) is not resolved in $I_T^{ch}$. Imaging at constant height ($z_1$) gives spin contrast in the $I_T^{ch}$ image comparable to SP-STM images at $z_0$ (Supplementary Fig. 2 and Supplementary Note 2), but does reveal more information on the magnetism within the spin spiral.

In contrast, the simultaneously probed $\varDelta f^{ch}$ resolves the atomic lattice for both the $[1\bar{1}0]$ and [001] direction. The $\varDelta f^{ch}$ image in Fig. 1(c) reflects the force gradient averaged within $z_1 \pm z_{mod}$ due to the oscillation of the tuning fork ($z_{mod} = 50$ pm). The total force that gives rise to the contrast in the image, is composed of chemical and magnetic exchange forces, and results in the observation of well-separated bright protrusions in the $\varDelta f^{ch}$ image, which we assign to the positions of the Mn atoms (see Supplementary Note 3 and Supplementary Fig. 3 for more information on the assignment of atomic sites). We note that electrostatic forces due to local charges are not expected to be significant because of the short screening length of the metallic substrate. Our extracted lattice constants ($\varDelta x_1 = (0.43 \pm 0.03)$ nm and $\varDelta x_2 = (0.30 \pm 0.03)$ nm agree well with the predicted ones[14]. We refer to the top sites, *t*, within each c(2 x 2) unit cell, *i* (Fig. 1(a)), within the spin spiral as $t_i$, where $i = 1$ and $i = 14$ correspond to $\varphi = 0°$ and $\varphi = 180°$ of the spin-spiral period (Fig. 1(d)), respectively. An analogous definition is used for the hollow sites $h_i$ (Fig. 1(a)). As the Mn layer is planar due to pseudomorphic growth on W(110)[14], atomic contrast emerging from chemical interaction is expected to result in the same variation of $\varDelta f^{ch}$ at all atomic sites ($t_i$ and $t'_i$) along the spin spiral. However, we observe a modulation $\Delta f_{ex,i} = \Delta f_i{}^{ch}(z_1) - \Delta f'_i{}^{ch}(z_1)$ of the contrast between $t_i$ and $t'_i$ along the [1$\bar{1}$0] direction (cf. solid vs. dashed circles at position $i = 13$ in Fig. 1(c)). We obtain $\Delta f_{ex,1}(z_1) = -(0.8 \pm 0.2)$ Hz, $\Delta f_{ex,7}(z_1) = (0.0 \pm 0.2)$ Hz and $\Delta f_{ex,14}(z_1) = (0.8 \pm 0.2)$ Hz, reflecting the reversal of the out-of-plane projection of the surface magnetization within the spin spiral as seen with SP-STM (see also sketch in Fig. 1(d)). We therefore assign $\Delta f_{ex,i}$ to variations of the magnetic exchange force along the spin spiral. We note that for different tips we have also observed an opposite correlation between $I_T{}^{ch}$ and $\Delta f^{ch}$ than that seen in Fig. 1(b) and Fig. 1(c). Our measurements show that the exchange force component of the measured force directly correlates with the out-of-plane projection of the surface magnetization. Therefore, this indicates that $\Delta f^{ch}$ directly probes not only the local antiferromagnetic order, but is also sensitive to the variations of the magnetic exchange interaction of the non-collinear magnetic moments on the surface resulting from the Dzyaloshinskii-Moriya interaction driven spin spiral.

**Quantifying the atomic-scale magnetic exchange force field.**

Next, we quantify the exchange force field at each atomic site as a function of tip-sample separation. We address both the top and hollow sites to study the influence of the local lattice configuration on the detected magnetic exchange interaction between each site and the tip, which we were not able to resolve in SP-STM mode. First, we focus on out-of-plane magnetic orientation of the atoms in the spin spiral with $p$ and $ap$ alignment (Fig. 1(d)). We record $\Delta f_i$ (from which we calculate $\Delta f_{ex,i}$) together with $I_{T,i}$ with respect to $\Delta z$ using the same atomic-scale tip (see Fig. 2(a) and the insets in (b) and (c), more details are available in the Supplementary Note 4 and Supplementary Fig. 4). The error margin reflects an uncertainty of ±0.2 Hz for $\Delta f_{ex,i}$ (Fig. 2(b,c), see Supplementary Note 2 for discussion of the influence of a surface tilt). Using the procedure described in Ref. [37], we derive the magnetic exchange forces $F_{ex,i}$ from $\Delta f_{ex,i}$ (Fig. 2(d,e)). As expected, we observe a sign reversal of the distance dependence when comparing $\Delta f_{ex,1}$ ($\Delta z$) and $\Delta f_{ex,14}(\Delta z)$, as well as for $F_{ex,1}(\Delta z)$ and $F_{ex,14}(\Delta z)$. Moreover, we observe a significant difference regarding the distance dependence of the exchange force field between top and hollow sites: for the hollow sites, the curves show a monotonous increase of the magnetic interaction (Fig. 2(c,e)). This is different from the curves acquired at the top sites (Fig. 2(b,d)) which show a change of the slope. The maximum absolute $\Delta f_{ex}$ values are ~ 3 Hz resulting in magnetic exchange forces up to 40 pN, *i.e.*, smaller by two orders of magnitude compared to the total forces (Supplementary Note 4 and Supplementary Fig. 4). However, the observed exchange forces here are larger by about a factor of three than probed on the non-collinear magnetic nano-skyrmion lattice of a monolayer Fe on Ir(111) using also a bulk Fe tip[27]. The distance-dependent variations of the exchange force field suggest that the measured exchange force results from different exchange mechanisms when comparing data for the top and hollow sites, as a result of the different nearest-neighbor configurations. Moreover, this evidence illustrates that $F_{ex,i}$ is highly sensitive to variations of the magnetic exchange force contributions on different atomic sites.

In order to have a measure for the distance-dependent spin-polarization, we consider the current asymmetry $A_i(\Delta z) = \frac{I_{T,i}(\Delta z) - I'_{T,i}(\Delta z)}{I_{T,i}(\Delta z) + I'_{T,i}(\Delta z)}$ with $I_{T,i}(\Delta z)$ acquired at $t_i$ or $h_i$ and $I'_{T,i}(\Delta z)$ acquired at $t'_i$ or $h'_i$. The error margins reflect an uncertainty of ±3% for $I_{T,i}$. The different signs of $A_1(\Delta z)$ and $A_{14}(\Delta z)$ (Fig. 2(f, g)) reflect the reverse relative alignments of the magnetic moments of the tip and Mn atoms (see insets in Fig. 2(b,c)). We observe that $A_1(\Delta z)$ and $A_{14}(\Delta z)$ do not show a significant variation in amplitude regardless of the atomic site (hollow or top) or distance. This is different from previous distance-dependent measurements of the Fe/Ir(111) skyrmion lattice[27], where $A(\Delta z)$ showed a monotonous increase. In addition, the absolute values of $A$ observed here are one order of magnitude smaller than for one monolayer Fe on Ir(111). Our findings here further show that detection of $F_{ex}(\Delta z)$, is more sensitive to atomic-scale variations, *i.e.* top and hollow sites, compared to $A(\Delta z)$, and provides more information about different exchange mechanisms.

So far, we have quantified the magnetic exchange force field at positions of the spin spiral with dominantly $p$ and $ap$ alignment (i.e., for $i = 1$ and 14). Next, we probe the exchange force field at different sites $i$ along the spin spiral using MExFIS, in order to quantify the variations of the exchange forces for the non-collinear orientations relative to the tip magnetization (Fig. 3a). We determine $F_{ex,i}(\Delta z)$ and $A_i(\Delta z)$ (Figs. 3(b,c)) at the hollow sites ($h_i$ and $h'_i$) at which we probe larger magnitudes of the exchange forces than on the top sites for the $p$ and $ap$ alignment (Fig. 2(e)). The gradual color change from blue to olive reflects the position within the spin spiral as sketched in Fig. 3(a). We observe a direct correlation between the magnetization of the spin spiral and the measured magnitude of the magnetic exchange force as shown in Fig. 3(b) for ($\lambda/4$) of the spin spiral. Given our error margins, we observe a gradual decrease of the absolute values of $F_{ex,i}(\Delta z)$ (as well as of $A_i(\Delta z)$) from $i = 1$ to $i = 7$. We assign the change of $F_{ex,i}(\Delta z)$ to the variation of the magnetic exchange interaction due to the different directions of the magnetic moments between $\varphi = 0°$ and $\varphi = 90°$ (Fig. 3(a)). We note that we also acquired data for $i = 7$ to $i = 14$ (see Supplementary Note 5, Supplementary Fig. 5 and Supplementary Fig. 6). We carefully acquired and analyzed all data to make sure that the tip has a dominantly magnetic out-of-plane sensitivity and that we can rule out measurement artifacts, e.g. piezo creep (for more information on the various types of force curves acquired, see Supplementary Note 6 and Supplementary Fig. 7). Nonetheless, we cannot exclude a misalignment of the tip magnetization by a few degrees from the out-of-plane direction and small changes of absolute distance due to a slight tilt of the sample (see Supplementary Note 5 for more details). Furthermore, we determined a spontaneous change of the tip magnetization during acquisition of the data for $i = 7$ to $i = 14$ (see Supplementary Fig. 6 for more information). Therefore, we have excluded this data from our discussion. Nevertheless, our results show that we can probe the variation of the magnetic exchange forces and current asymmetry along the spin spiral.

**First-principles calculations.**

In order to understand the experimental observations, we have performed first-principles calculations for Mn/W(110) and for the interaction with a magnetic tip (see Methods for computational details and Supplementary Note 7 and Supplementary Fig. 8 for the studied geometries). Although an Fe bulk tip was used in the experiments, the tip was prepared by gentle dipping into the Mn layer with co-adsorbed Co adatoms. Therefore, we consider the influence of three different tip terminations (Fe, Co, Mn) of Fe-based tips. In our calculations we approached the tip on both the top and the hollow site of the Mn monolayer (cf. Fig. 1(a)) and considered a parallel ($p$) and an antiparallel ($ap$) alignment of the magnetic moments between the tip apex atom and the Mn surface atom (top site) or nearest-neighbor Mn surface atoms (hollow site). The effect of structural relaxations due to tip-sample interaction does not change our conclusions, as shown for the Fe tip (see Supplementary Note 8 and Supplementary Fig. 9). The magnetic exchange energy, $E_{ex}(d)$, can be calculated from the total energies of the $ap$- and $p$- configurations:

$$E_{ex}(d) = E_{ap}(d) - E_p(d), \qquad (1)$$

where $E_{ap}(d)$ and $E_p(d)$ denotes the total energies of the $ap$- and $p$- configurations, respectively. Therefore, $E_{ex}(d) > 0$ and $E_{ex}(d) < 0$ indicate ferromagnetic and antiferromagnetic coupling, respectively. Similarly, the magnetic exchange force $F_{ex}(d)$ is defined as:

$$F_{ex}(d) = F_{ap}(d) - F_p(d). \qquad (2)$$

Here, $F_{\mathrm{ap}}(d)$ and $F_{\mathrm{p}}(d)$ represent the total forces on the tip along the [110] (i.e., the $z$-) direction of Mn/W(110) for $ap$- and $p$- configurations, respectively.

Fig. 4 shows the calculated magnetic exchange energies $E_{\mathrm{ex}}$ for top (a,c) and hollow sites (e,g) and the corresponding exchange forces $F_{\mathrm{ex}}$ (b,d,f,h) as a function of $\Delta d$ that is related to the experimentally determined $\Delta z$ by an offset as defined further below. $\Delta d$ is defined as $d$–$d_0$, where $d_0 = 0.5$ nm. Already at first glance, we observe that the exchange energy and forces qualitatively differ for the two sites on the Mn monolayer – in agreement with the experimental observation – as well as for different tip terminations. For Mn-terminated tips, we find a regime of ferromagnetic coupling on the top site at large distances, i.e. $E_{\mathrm{ex}}(\Delta d) > 0$ for $\Delta d >$ -0.16 nm (Fig. 4(a)). For smaller $\Delta d$, the coupling is antiferromagnetic. In contrast, for a Fe- or Co-terminated tip, only antiferromagnetic coupling is observed (Fig. 4(c)). We attribute the change of sign for Mn-terminated tips to a transition from an indirect long-range ferromagnetic exchange mechanism between the Mn $d$-states mediated by $s$ electrons (Zener-type exchange), to a short-range direct antiferromagnetic exchange between $d$-orbitals of Mn apex and surface atom (as suggested in Ref. [31]). The Zener-type exchange mechanism is much weaker for Fe and Co due to their smaller $s$-$d$ coupling[38, 39, 40], and no transition occurs, in agreement with a previous work[30]. The different types of exchange mechanisms can be seen in spin-resolved charge-density difference plots (see Supplementary Note 9 and Supplementary Fig. 10). For hollow sites, the coupling is always ferromagnetic for the Mn-terminated tip (Fig. 4(e)). For Fe- and Co-terminated tips, antiferromagnetic coupling at large tip-sample separations changes to a ferromagnetic coupling at small tip-sample separations (Fig. 4(g)). The relatively small values of $E_{\mathrm{ex}}$ and $F_{\mathrm{ex}}$ for hollow sites and the change of sign for Fe- and Co-terminated tips can be attributed to competing exchange interactions between the tip apex atom and nearest neighbor and next-nearest neighbor surface Mn atoms (cf. Fig. 1(a)). For Mn-terminated tips the Zener-type exchange dominates. Our calculations reveal that the magnetic exchange force is highly sensitive to competing magnetic exchange interactions between the tip and the Mn layer depending on the chemical tip termination as well as on the atomic site.

The calculated exchange force curves for a Mn-terminated tip (Fig. 4(f)) agree qualitatively with the experimental data shown in Fig. 2(d,e) while Fe- and Co-terminated tips (Fig. 4(h)) cannot explain the observations. From a comparison between the onset of magnetic exchange forces in the calculations and in the experimental curves we estimate that $\Delta d = 0.0$ nm corresponds to a tip displacement of about $\Delta z = -0.16$ nm. This estimate is also in reasonable agreement regarding the total measured and calculated forces (see Supplementary Fig. 3 and Supplementary Fig. 4). For the discussion of the calculated and experimental $F_{ex}$, we focus on the tip-sample separations between $\Delta d = -0.15$ nm and $\Delta d = 0.0$ nm corresponding to $\Delta z = -0.31$ nm and $\Delta z = -0.16$ nm, respectively. The experimental exchange force $F_{ex}(\Delta z)$ on top sites first rises up to an absolute maximum value and changes its sign when the tip is brought closer to the surface (Fig. 2(d)). At the hollow sites, $F_{ex}(\Delta z)$ exhibits a monotonous increase of the absolute values (Fig. 2(e)). This agrees with the change in the calculations from positive exchange force for $\Delta d > -0.1$ nm to negative exchange force for $\Delta d < -0.1$ nm on the top site for the Mn-terminated tip (Fig. 4(b)), while there is a monotonous rise of the exchange force on the hollow sites (Fig. 4(f)). We point out that we have observed many different types of distance-dependent curves experimentally (see Supplementary Note 10 and Supplementary Fig. 11), which can result from a combination of different tip terminations and tip geometries, as well as the presence of irreversible dissipative mechanisms such as surface-induced magnetization flipping of the tip or hydrogen diffusion (see

Supplementary Note 6 and Supplementary Fig. 7). For our discussion here, we excluded all experimental data showing any indications of irreversible dissipative processes.

To understand the experimentally observed small and site-independent spin polarization, we analyze the vacuum local density of states (LDOS) at the top and hollow sites. Within the Tersoff-Hamann model of STM[41, 42], the differential conductance is proportional to the LDOS a few Å above the surface, and the $A(\varDelta z)$ curves in spin-polarized measurements (Fig. 2(f,g)) can be compared to the spin polarization of the vacuum LDOS[43, 44]. As shown in the Supplementary Fig. 12, there is very little difference between the spin polarization on the hollow and the top site, in agreement with the experimental observation. In addition, we find quite a small value on the order of 20% around $E_F$. As shown in the orbitally decomposed LDOS (see Supplementary Fig. 13 and Supplementary Note 11) the small spin polarization is due to $p_z$ states which dominate the vacuum LDOS. Typically, the spin-polarized LDOS on 3$d$ transition-metal films and surfaces is dominated by $d$ states with large values of the spin polarization on the order of 50 to 80%[43, 45]. However, since Mn exhibits a very large exchange splitting and a large magnetic moment of 3.6 $\mu_B$ the $d$ bands are far from the Fermi energy and $p$ states dominate which are much less spin polarized. This is a rather different finding than for the recently studied Fe monolayer on Ir(111), where the magnetic moment is significantly smaller and the $d$-states strongly contribute to the spin polarization close to the Fermi energy[18].

**Discussion.**

In conclusion, we quantify the atomic-scale variation of the magnetic exchange force field between a ferromagnetic tip and a chiral magnetic surface. Using constant-height SPEX imaging, we spatially resolve the magnetic exchange interaction of the tip and the cycloidal spin spiral in one monolayer Mn on the W(110) surface. While the constant-height current image shows the expected magnetic contrast, but only resolves the atomic lattice in one direction, the constant-height frequency shift image resolves the magnetism within the entire antiferromagnetic c(2 x 2) unit cell. Using MExFIS, we quantify the exchange force field along the spin spiral with atomic resolution and observe a strong influence of the exchange forces which depend on the local spin configurations on hollow and top sites, indicating different exchange regimes. In contrast, the site-dependent variations are absent in the current, reflecting that magnetic exchange forces are more sensitive to changes of the atomic environment. Utilizing first-principles calculations, we understand those site-dependent differences of the exchange forces and quantify the chemical contribution of the tip apex to the overall distance dependence of the force field and the various possible exchange pathways. The experimental findings qualitatively agree with the exchange interactions probed by a Mn-terminated tip that reveals an interplay between antiferromagnetic and Zener-type ferromagnetic exchange mechanisms. The experimentally observed, surprisingly weak spin polarization is explained by the dominant contribution of $p$-states near $E_F$ due to the large exchange splitting in Mn.

Our work provides the first combined experimental and theoretical study to quantify the different exchange force regimes of a chiral magnetic structure on the atomic scale. As a future perspective, SPEX can be used to simultaneously quantify the influence of spin-orbit coupling on both the current and the magnetic interactions in chiral magnetic structures, as well as characterize them with antiferromagnetic probes or functionalized tips. Furthermore, 3D-MExFIS opens a promising route to quantify the various direct and indirect exchange forces, with atomic-scale resolution, in addition to extracting dissipation mechanisms, providing direct access to interaction energies in comparison to state-of-the-art methods based solely on tunneling current[7, 46].

**Methods**

**Experimental procedures.**

We performed SPEX imaging using a bulk Fe tip with a magnetization out of plane to the surface (see Supplementary Note 1 for more details) in a modified commercial ultra-high vacuum STM/AFM[47] operating at a base temperature of $T$ = 5.8 K. Force detection was done in non-contact frequency-modulation mode, utilizing a tuning fork-based qPlus sensor[48] with the free prong oscillating at its resonance frequency ($f_0$ = 30.8 kHz) and with a quality factor larger than 25000. The bias voltage $V_s$ was applied to the sample, and we performed both constant-current (cc) and constant-height (ch) imaging. The W(110) surface was cleaned by annealing ($T \sim 1800$ K) in an oxygen atmosphere ($p \sim 3$ x $10^{-7}$ mbar) followed by a final flash to $T \sim 2800$ K. The 0.8 monolayer of Mn were deposited from Mn granules (99.99% purity, MaTeck) by using an e-beam evaporator. The W(110) was kept at room temperature during deposition, and was annealed ($T \sim 800$ K) after the Mn deposition. To characterize our magnetic tips, we utilize the spin-direction dependent appearance of co-adsorbed Co adatoms as a reference[49, 50], which were deposited onto the cold sample ($T \sim 6$ K) from a Co rod source (99.99% purity, MaTeck) by using an e-beam evaporator, with 1% of coverage (see Supplementary Note 1 for more details). Tips with spin contrast were obtained by gently dipping a bulk Fe tip into the Mn/W(110) monolayer. The tip was brought closer to the Mn layer by values between -600 pm to -900 pm with respect to the tunneling position $z_0$, which is defined by the constant-current feedback setpoint $V_s$ = -10 mV and $I_T^{cc}$ = -2 nA. We used the $\Delta f$ values at this feedback setpoint to characterize the bluntness of our tips, which were typically between -10 Hz and -17 Hz.

**Density functional theory.**

We applied first-principles calculations using a plane wave-based density functional theory (DFT) code VASP[51, 52] within the projector augmented wave method (PAW)[53, 54]. We used the generalized gradient approximation (GGA) of Perdew-Burke-Ernzerhof (PBE) for the exchange correlation[55, 56]. The combined system of tip and sample was calculated in a supercell geometry (see Supplementary Fig. 8). The Mn/W(110) surface was modelled using by a symmetric slab with 5 W layers and 1 layer of Mn on each side. The local magnetic order of the Mn/W(110) surface can be approximated as collinear antiferromagnetic (AFM) due to the long periodicity of the spin spiral ground state. We have used a c(6×6) AFM surface unit cell with the GGA lattice constant of W (3.17 Å), which is in good agreement with the experimental value of 3.165 Å. The tip was modelled by a 14-atom pyramid in a bcc(001) orientation (see Supplementary Fig. 8). We have used ferromagnetic Fe to model the tip. For the alloyed tips, we have used Co and Mn atoms as the apex atom. For a Mn-terminated tip apex, antiferromagnetic coupling occurs between the Mn and Fe atoms. The tip and the surface were initially relaxed independently before considering the coupled system, *i.e.*, the tip-sample interaction. For the isolated tips, i.e., without any tip-sample interaction, the apex atom and the four atoms of the adjacent layer have been relaxed. The in-plane interatomic distances between the base atoms are kept constant at the GGA lattice constant of Fe (2.85 Å). Note that for Fe tips we have also considered the effect of structural relaxations due to tip-sample interaction (see Supplementary Note 8). We have used a 450 eV energy as plane wave basis set cutoff. A $(5 \times 5 \times 1)$ $\bar{\Gamma}$ centered **k**-grid mesh was used for the Brillouin zone integration. For every tip-sample separation $d$ (defined as the distance between the centers of the tip apex atom and the surface Mn atom underneath without structural relaxations due to the tip-sample interaction), we carried out two sets of calculation for $p$ and *ap* alignment of the magnetic moments between the tip apex atom and the Mn surface atom (top site) or nearest-neighbor Mn surface atoms (hollow site). The magnetic exchange energy was calculated from the total energies in the *p* and *ap* alignment. The forces were calculated by using the Hellmann-Feynman theorem[57].

**Acknowledgements**

We acknowledge financial support from the NWO VIDI project: 'Manipulating the interplay between superconductivity and chiral magnetism at the single atom level' with project number 680-47-534 and the NWO Physics/f grant (project: 680.91.003 SPEX). This project has received funding from the European Research Council (ERC) under the European Union's Horizon 2020 research and innovation programme (SPINAPSE, grant agreement No 818399). S.Ha. and S.He. acknowledge financial support from the Deutsche Forschungsgemeinschaft via SFB677. S.Ha. M.G. and S.He. thank the Norddeutscher Verbund für Hoch-und Höchleistungsrechnen (HLRN) for providing computational resources. W.J. acknowledges support from the Alexander von Humboldt Foundation via the Feodor Lynen Research Fellowship.

**Author contributions**

N.H., S.He. and A.A.K. designed the experiment. T.-C.H., W.J. and N.H. carried out the measurements and did the analysis of the experimental data. M. G. and S. Ha. performed the DFT calculations for the clean Mn/W(110) surface. S. Ha. performed the DFT calculations for the interaction of the tip with the Mn/W(110) surface. S. Ha. and S. He. analyzed the results of the DFT calculations. N.H., T.-C.H., S.Ha., S.He., D.W. and A.A.K. wrote the manuscript. All authors contributed to the discussion and interpretation of the results as well as the discussion of the manuscript.

**Additional information**

Supplementary Information accompanies this paper at …

## Data availability

The authors declare that all relevant data are included in the paper and its Supplementary

Information file. Additional data are available from the corresponding author upon reasonable request.

## Competing interests

The authors declare no competing interests.

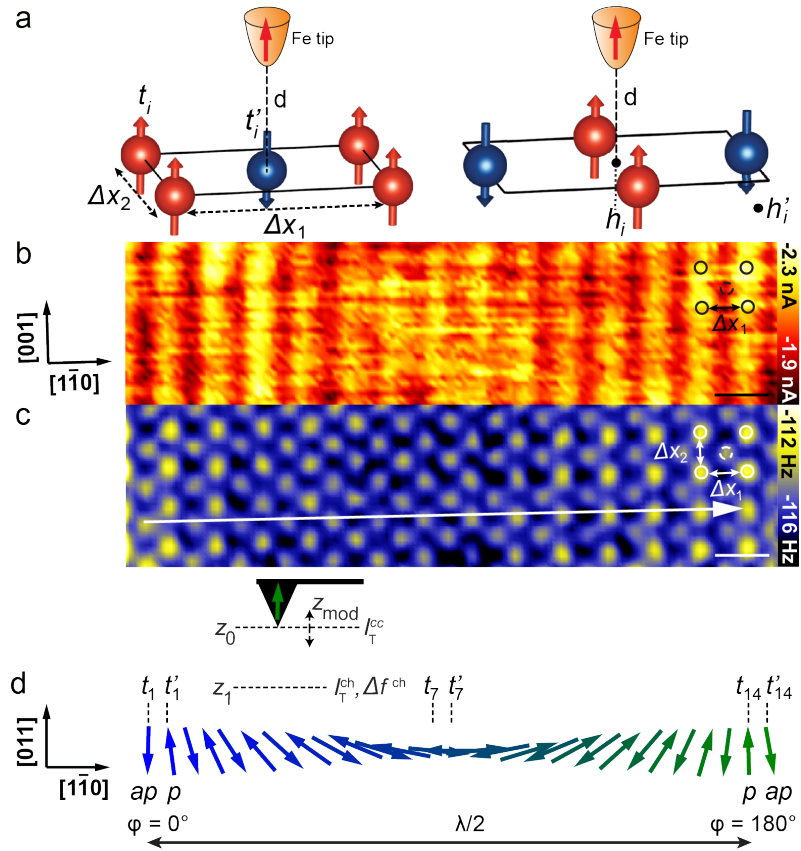

**Figure 1: Atomic-resolution SPEX imaging of the Mn spin spiral.** (a) Sketches of the antiferromagnetic c(2 x 2) unit cell which locally approximates the spin spiral. The extracted lattice constants from the experiment ($\Delta x_1$ and $\Delta x_2$), the tip-Mn distance $d$, the different top ($t_i$ and $t'_i$) and hollow ($h_i$ and $h'_i$) sites are indicated for the tip positioned at $t'_i$ (left) and $h_i$ (right). (b) Current $I_T^{ch}$ and (c) frequency shift $\Delta f^{ch}$ images of one half of the spin-spiral period (see (d), both images line-flattened, frequency shift image Gaussian-smoothed by two points) measured at constant height ($z_1$) that is by 0.29 nm closer to the surface than the height $z_0$ at which the current feedback loop was opened ($V_s$ = -10 mV and $I_T^{cc}$ = -2 nA). Parameters: oscillation amplitude $z_{mod}$ = 50 pm, $V_s$ = -0.1 mV, tip magnetization normal to the surface. The arrow in (c) depicts the contrast variation due to the reversal of a single spin along the [1$\bar{1}$0] direction. The scale bars in (c) and (d) correspond to 0.5 nm. (d) Side-view sketch of one half of the cycloidal spin spiral along the [1$\bar{1}$0] direction in one monolayer Mn on W(110) together with the experimental measurement scheme. The magnetic moments of the spin spiral have a parallel (*p*) alignment with respect to the tip magnetization for $t'_1$ and $t_{14}$, and an anti-parallel (*ap*) alignment for $t_1$ and $t'_{14}$. Positions $t_7$ and $t'_7$ indicate the position with in-plane magnetization.

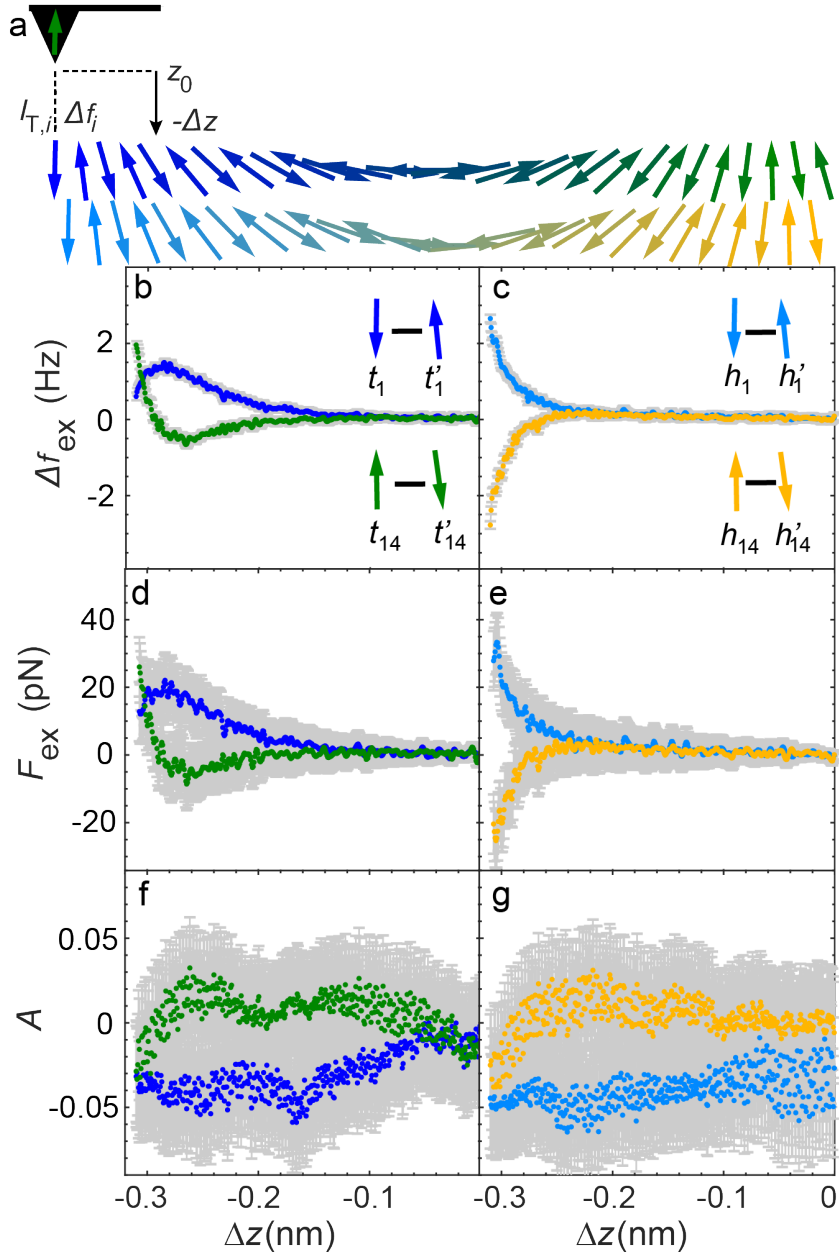

**Figure 2: Quantifying the atomic-scale magnetic exchange force field at top and hollow sites.** (a) Sketch of the distance-dependent measurement scheme for the top and hollow sites along the spin spiral. The upper and lower sketch of the spiral reflects the color code for top and hollow sites, respectively. Data is acquired at top ($t_1$, $t'_1$ and $t_{14}$, $t'_{14}$) and hollow ($h_1$, $h'_1$ and $h_{14}$, $h'_{14}$) sites within the c(2 x 2) unit cell (Fig. 1(a)). (b,c) Frequency shift difference $\Delta f_{ex}$ versus tip displacement $\Delta z$, reflecting the magnetic exchange force contribution ($\Delta f_{ex}(\Delta z) = \Delta f(\Delta z) - \Delta f'(\Delta z)$). (d, e) Magnetic exchange force $F_{ex}(\Delta z)$ derived from $\Delta f_{ex}(\Delta z)$. (f,g) Current asymmetry $A(\Delta z)$. The grey error margins reflect an uncertainty of ±0.2 Hz and ±3% for $\Delta f_{ex}$ and $I_T$, respectively. All curves are smoothed by three points using a Savitzky-Golay filter. Prior to calculation of $F_{ex}$, $\Delta f_{ex}$ has been smoothed by five points. Parameters: oscillation amplitude $z_{mod} = 50$ pm, $V_s = -0.05$ mV, out-of-plane tip magnetization. The position $\Delta z = 0$ nm corresponds to our reference height $z_0$, that is defined by a setpoint of the current feedback loop of $V_s = -10$ mV; $I_T^{cc} = -2$ nA. Negative $\Delta z$ reflects smaller tip-surface distances.

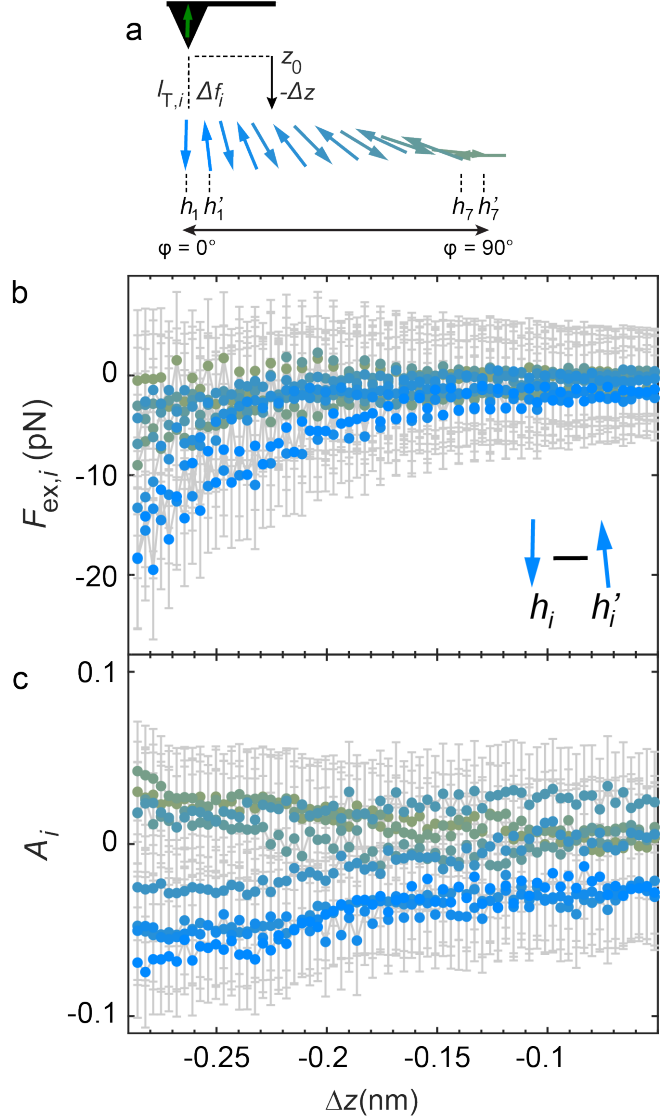

**Figure 3: (a) Quantifying the atomic-scale magnetic exchange force field along the spin spiral.** Sketch of the cycloidal spin spiral for the hollow sites and illustration of the distance-dependent measurement scheme. (b) Magnetic exchange force $F_{ex,i}(\Delta z)$ derived from $\Delta f_{ex,i}(\Delta z)$, versus the tip displacement $\Delta z$ for different hollow sites $h_i/h'_i$ (with $i = 1$ to $i = 7$). (c) Current asymmetry $A_i(\Delta z)$. The grey error margins reflect an uncertainty of $\pm 0.2$ Hz and $\pm 3\%$ for $\Delta f_i/\Delta f'_i$ and $I_{T,i}/I'_{T,i}$, respectively. Parameters: oscillation amplitude $z_{mod} = 43$ pm, $V_s = -0.1$ mV. During the data acquisition, the tip magnetization may differ from the out-of-plane orientation which was characterized prior to and after the measurements. The position $\Delta z = 0$ nm corresponds to our reference height $z_0$, that is defined by a setpoint of the current feedback loop of $V_s = -10$ mV; $I_T^{cc} = -2$ nA. Negative $\Delta z$ reflects smaller tip-surface distances.

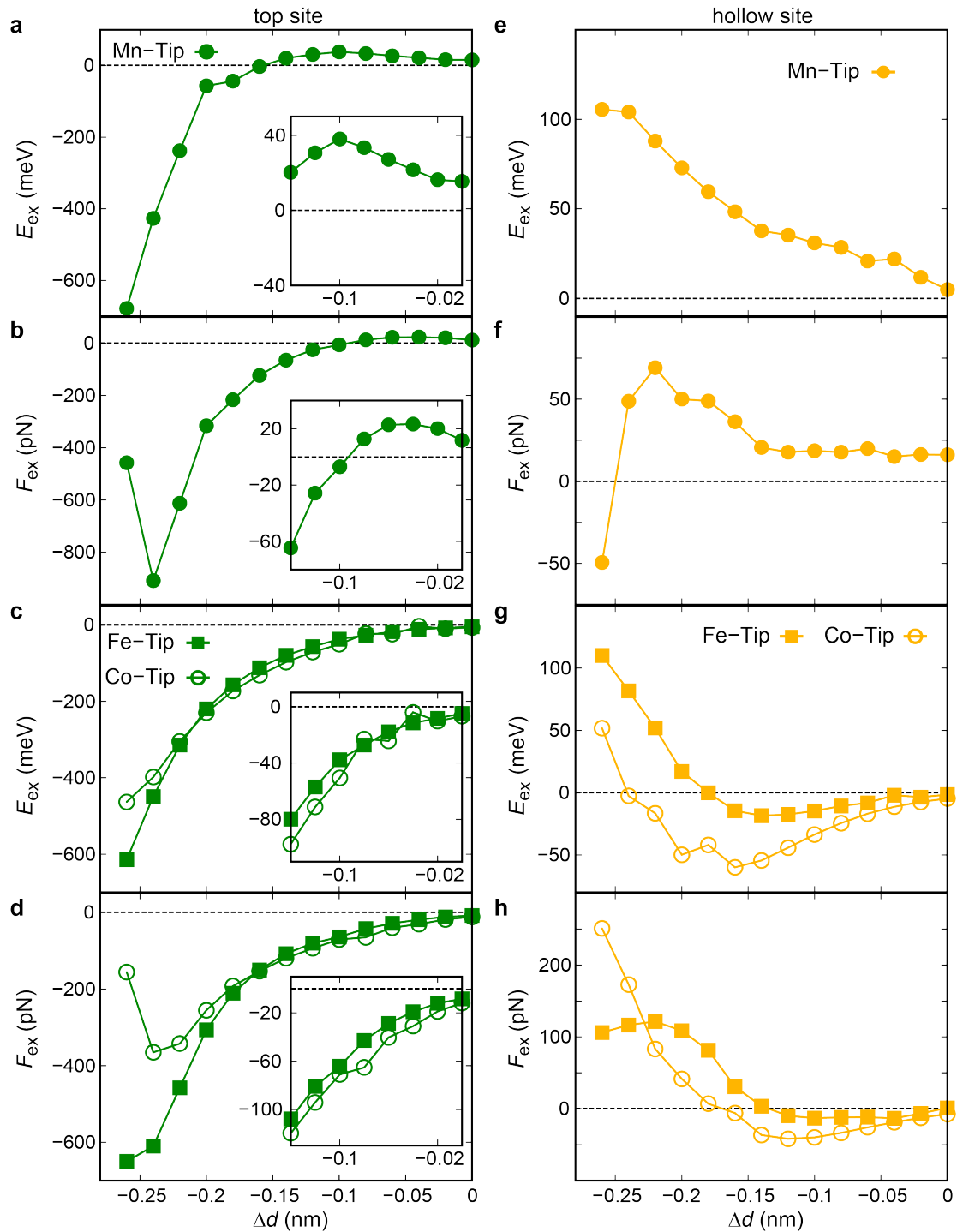

**Figure 4: First-principles calculations of the exchange interaction**. Calculated magnetic exchange energies $E_{ex}$ on top (a,c) and hollow (e, g) sites for collinear alignments of the magnetic moments of tip and Mn atoms, for Fe-, Co-terminated (c, g), and Mn-terminated (a, c) tips versus $\Delta d$ that is defined as $d–d_0$, where $d_0 = 0.5$ nm and $d$ is the tip-sample separation. Corresponding exchange forces $F_{ex}$ on top (b, d) and hollow sites (f, h) versus $\Delta d$.